\documentclass[twocolumn,preprintnumbers, amssymb,amsmath,aps,floatfix,prd,nofootinbib,superscriptaddress,showpacs]{revtex4-1}

\usepackage{epsfig}
\usepackage{bm}
\usepackage{amssymb}
\usepackage{amsmath}
\usepackage{color}
\usepackage{subfigure}
\usepackage[colorlinks,
            linkcolor=blue,
            anchorcolor=black,
            citecolor=blue
            ]{hyperref}

\newcommand{\beq}{\begin{eqnarray}}
\newcommand{\eeq}{\end{eqnarray}}

\newcommand{\nn}{\nonumber \\}

\begin{document}
\title{Semi-inclusive Diffractive Deep Inelastic Scattering at Small-$x$ }

\author{Yoshitaka Hatta}
\affiliation{Physics Department, Building 510A, Brookhaven National Laboratory, Upton, NY 11973, USA}
\affiliation{RIKEN BNL Research Center,  Brookhaven National Laboratory, Upton, New York 11973, USA}

\author{Bo-Wen Xiao}
\affiliation{School of Science and Engineering, The Chinese University of Hong Kong, Shenzhen 518172, China}

\author{Feng Yuan}
\affiliation{Nuclear Science Division, Lawrence Berkeley National
Laboratory, Berkeley, CA 94720, USA}


\begin{abstract}
Inspired by a recent study by Iancu, Mueller, and Triantafyllopoulos~\cite{Iancu:2021rup}, we propose semi-inclusive diffractive deep inelastic scattering (SIDDIS) to investigate the gluon tomography in the nucleon and nuclei at small-$x$. The relevant diffractive quark and gluon parton distribution functions (DPDF) can be computed in terms of the color dipole S-matrices in the fundamental and adjoint representations, respectively. 
\end{abstract}
\maketitle

\section{Introduction}

Nucleon tomography in terms of various `multi-dimensional' parton distribution functions is one of the ultimate goals of the current and future facilities in high energy nuclear and particle physics~\cite{Boer:2011fh, AbelleiraFernandez:2012cc, Accardi:2012qut}. 
These include the transverse momentum dependent distributions (TMDs) and the generalized parton distributions (GPDs), which provide 
different perspectives of the internal structure of hadrons and nuclei. The so-called quantum phase space Wigner distributions~\cite{Ji:2003ak, Belitsky:2003nz} of partons are regarded as the mother distributions since they ingeniously encode the complete information about how partons are distributed both in position and momentum spaces. 

At small-$x$ in the gluon saturation regime, the gluon Wigner distribution is intimately connected to the well-known color dipole S-matrix in the Color-Glass-Condensate (CGC) formalism~\cite{Mueller:1993rr,Mueller:1999wm,McLerran:1993ni,McLerran:1993ka,McLerran:1994vd} which  has been a subject of intensive study in the last few decades~\cite{Gelis:2010nm, Kovchegov:2012mbw}. In Ref.~\cite{Hatta:2016dxp}, it was suggested that the diffractive dijet production in $ep$/$eA$ collisions~\cite{Hatta:2016dxp,Altinoluk:2015dpi,Zhou:2016rnt,Hagiwara:2017fye,Mantysaari:2019csc,Mantysaari:2019hkq} may provide a direct probe of the gluon Wigner distribution. Recently, Iancu, Mueller, and Triantafyllopoulos \cite{Iancu:2021rup} have considered the correction of an additional semi-hard gluon radiation to this process, or `trijet' production. A remarkable feature 
is that the leading dijet can have a much larger transverse momentum than the saturation momentum $Q_s$, yet, the process is still sensitive to gluon saturation due to the third jet, whose transverse momentum is of the order of $Q_s$. 
Another remarkable feature is that the calculated cross section factorizes in terms of the gluon 
PDF of the `Pomeron', or equivalently, as we shall see, the gluon diffractive parton distribution function (DPDF). The DPDFs are important ingredients for the QCD factorization of diffractive hard processes~\cite{Collins:1997sr, Berera:1995fj,Trentadue:1993ka}, see, a recent phenomenology study~\cite{Armesto:2019gxy} and reference therein. 
At small-$x$, the DPDFs are connected to the color dipole S-matrix~\cite{Hebecker:1997gp,Buchmuller:1998jv,Golec-Biernat:1999qor,Hautmann:1999ui,Hautmann:2000pw,Golec-Biernat:2001gyl}, and are therefore systematically calculable including the gluon saturation effects. 

\begin{figure}[t]
\begin{center}
\includegraphics[width=0.3\textwidth]{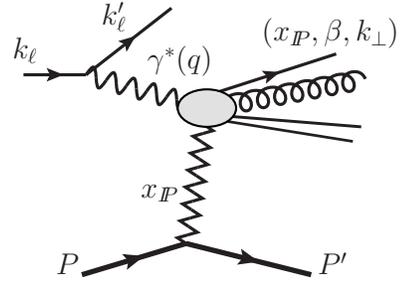}
\end{center}
\caption[*]{Semi-inclusive diffractive DIS process, where the diffractive parton distributions of quark and gluon can be measured through final state particles, including hadron/jet productions. An incoming proton with momentum $P$ diffractively scatters into the final state nucleon with momentum $P'$ and deposits a longitudinal momentum fraction of $x_{I\!\!P}$ into hard interaction with the virtual photon. The usual Bjorken $x$ is defined as $x_B=\beta x_{I\!\!P}$ for inclusive diffractive DIS. }
\label{fig:siddis}
\end{figure}

Following these developments, in this paper, we will further demonstrate that the semi-inclusive diffractive DIS (SIDDIS), see Fig.~\ref{fig:siddis}, can provide a unique perspective of gluon tomography at small-$x$, where the quark and gluon DPDFs can be systematically computed from the operator definitions consistent with the QCD factorization~\cite{Collins:1997sr} in the dipole formalism. This opens up new opportunities to investigate the gluon Wigner distribution and gluon saturation at the future electron-ion collider (EIC). More importantly, the QCD factorization results in terms of DPDFs are consistent with the CGC calculations in the kinematics that both apply.

Compared to the usual semi-inclusive DIS (SIDIS)~\cite{Mulders:1995dh,Boer:1997nt,Bacchetta:2006tn}, the diffractive process requires a large rapidity gap, $Y_{I\!\!P}\sim \ln(1/x_{I\!\!P})\gg 1$ with small $x_{I\!\!P}$ considered here, between the nucleon remnant and the hard interaction part, which can be easily identified in experiment. Combining the methods applied in the QCD factorization for SIDIS (see, e.g., \cite{Ji:2004wu,Collins:2011zzd}) and hard diffractive processes~\cite{Collins:1997sr}, we expect that the QCD factorization for the SIDDIS is also valid, and we can safely extract the relevant DPDFs from the experiment. For the transverse momentum dependent observables, we will need the TMD fragmentation function and the associated soft factor as well, which can be defined accordingly. Integrating over the transverse momentum will lead to a collinear factorization for SIDDIS, where the soft factor does not contribute.    

The study of hard diffractive processes has a long history. Theoretically, there are three different approaches based on QCD factorization framework. For inclusive hard diffractive process, there is a collinear factorization in terms of the DPDFs. In the CGC approach, all diffractive processes can be described by the dipole scattering amplitudes~\cite{Golec-Biernat:1999qor,Hatta:2006hs,Marquet:2007nf,Kowalski:2008sa}. Meanwhile, there is also a generalized parton distribution (GPD)~\cite{Ji:1996ek,Mueller:1998fv,Ji:1996nm} approach to describe the hard exclusive processes~\cite{Ji:1996ek,Mueller:1998fv,Ji:1996nm,Radyushkin:1997ki,Goeke:2001tz,Diehl:2003ny,Belitsky:2005qn,Boffi:2007yc,Boer:2011fh,Accardi:2012qut}, including deeply virtual Compton scattering (DVCS). The consistency between the collinear GPD formalism and the CGC/dipole formalism has been shown for the DVCS process at small-$x$~\cite{Hatta:2017cte}. The investigation of SIDDIS in this paper extends this consistency and provides a unified method that connects all the above-mentioned approaches. Our discussions below are limited to the so-called coherent diffractive processes. But, this can be extended to the in-coherent diffractive process as well.


\section{Diffractive PDFs from Dipole Amplitude at Small-$x$}

It has been shown that, at small-$x$, the quark and gluon TMD distribution functions are directly related to the color dipole S-matrix  in the CGC formalism~\cite{McLerran:1993ni,McLerran:1993ka,McLerran:1998nk,Mueller:1999wm,Marquet:2009ca,Dominguez:2010xd,Dominguez:2011wm,Belitsky:2002sm,Xiao:2010sa,Xiao:2017yya}. 
In this section, we use the same method to establish the connection between the DPDFs and the color dipole. The result in the gluon case is equivalent, up to the normalization factor, to the `unintegrated gluon distribution of the Pomeron' calculated in Ref.~\cite{Iancu:2021rup}.  

Let us begin with the standard definition of the quark DPDF \cite{Berera:1995fj} generalized to include the transverse momentum ($k_\perp$) dependence  
 \begin{eqnarray}
&&2E_{P'}\frac{df_q^{D}(x,k_\perp;x_{I\!\!P},t)}{d^3P'}=\int
        \frac{d\xi^-d^2\xi_\perp}{2(2\pi)^6}e^{-ix\xi^-P^++i\vec{\xi}_\perp\cdot
        \vec{k}_\perp} \nonumber\\
        &&~\times \langle
PS|\overline\psi(\xi){\cal L}_{n}^\dagger(\xi)\gamma^+|P'X\rangle \langle P'X|{\cal L}_{n}(0)
        \psi(0)|PS\rangle  \ ,\label{tmdun}
\end{eqnarray}
where the future pointing gauge link in the fundamental representation of QCD is defined as $ {\cal L}_{n}(\xi) \equiv \exp\left(-ig\int^{\infty}_0 d\lambda \, v\cdot A(\lambda n +\xi)\right)$. Here, $n$ represents a light-cone vector conjugate to the nucleon momentum $n^2=0$ and $n\cdot P=1$. The final state nucleon carries momentum $P'=P+\Delta$ with $t=\Delta^2$.  $x_{I\!\!P}= n\cdot (P-P')$ is the momentum fraction of the incoming nucleon carried by the Pomeron. We introduce the momentum fraction of the Pomeron carried by the quark $\beta=x/x_{I\!\!P}$. Integrating over $k_\perp$, we recover the collinear quark DPDF. 

Just like usual TMDs, the naive definition (\ref{tmdun}) contains end-point singularities at higher orders which will be cured by the soft factor subtraction. This will introduce the associated TMD-like evolution and resummation~\cite{Collins:2011zzd}, which are important for phenomenology applications. We will come back to this issue in the future. In the following, we will neglect such higher order effects. We also mention that (\ref{tmdun}) is similar to, but different from the generalized transverse momentum dependent parton distribution (GTMD)~\cite{Meissner:2009ww} or the  Wigner distribution~\cite{Belitsky:2003nz}. In particular, the GTMD appears in the amplitude for the exclusive process, whereas the TMD DPDF appears in the cross section for semi-inclusive diffractive processes. However, there exist strong connections between them, as we will show  below.

Similarly, we can define the gluon TMD DPDF
 \begin{eqnarray}
&&2E_{P'}\frac{df_g^{D}(x,k_\perp;x_{I\!\!P},t)}{d^3P'}=\int
        \frac{d\xi^-d^2\xi_\perp}{xP^+(2\pi)^6}e^{-ix\xi^-P^++i\vec{\xi}_\perp\cdot
        \vec{k}_\perp} \nonumber\\
        &&~\times\!\! \langle
PS|F^{+\mu}(\xi){\cal L}_{n}^\dagger(\xi)\gamma^+|P'X\rangle \!\langle P'X|{\cal L}_{n}(0)
        F_\mu^{\, +}(0)|PS\rangle  \ ,\label{tmdung}
\end{eqnarray}
where the gauge link is in the adjoint representation.  

\begin{figure}[t]
\begin{center}
\includegraphics[width=0.35\textwidth]{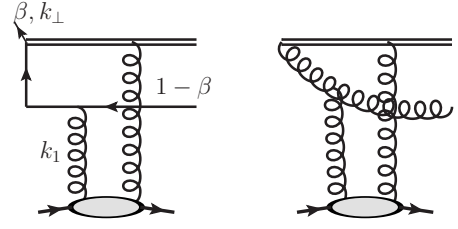}
\end{center}
\caption[*]{The typical Feynman diagrams for the Diffractive quark (left) and gluon (right) distribution functions computed from the dipole amplitudes at small-$x$. }
\label{fig:dff}
\end{figure}

The main goal of our paper is to compute the quark and gluon DPDFs in the CGC framework. The typical Feynman diagrams are shown in Fig.~\ref{fig:dff}, where the double lines represent gauge links. To the order of accuracy, the final state $|X\rangle$  is saturated by a single antiquark and a gluon  for the quark and gluon DPDFs, respectively.  Calculating the left diagram of Fig.~\ref{fig:dff} and its complex conjugate,  we find the following result,
\begin{eqnarray}
&&x\frac{d\, f_q^D(\beta,k_\perp;x_{I\!\!P})}{dY_{I\!\!P} dt} =   \int
d^2k_{1\perp}d^2k_{2\perp} {\cal F}_{x_{I\!\!P}}(k_{1\perp},\Delta_\perp)
\nonumber \\ 
&&~~\times{\cal F}_{x_{I\!\!P}}(k_{2\perp},\Delta_\perp)
\frac{N_c\beta}{2\pi} {\cal T}_q(k_\perp,k_{1\perp},k_{2\perp}) \ ,
\label{dffquark}
\end{eqnarray} 
for the quark DPDF at small-$x$. Here, $k_1$ represents the momentum for one of the two vertical gluon lines in the left diagram of Fig.~\ref{fig:dff}. For the complex conjugate of this diagram, we use $k_2$. ${\cal T}_q$ represents the sum of  four terms  ${\cal T}_q\equiv T_q(k_\perp,k_{1\perp},k_{2\perp})-T_q(k_\perp,0,k_{2\perp})-T_q(k_\perp,k_{1\perp},0)+T_q(k_\perp,0,0)$ where 
\begin{eqnarray}
&&    T_q(k_\perp,k_{1\perp},k_{2\perp})=\nonumber\\
&&\frac{ k_{1\perp}'\cdot k_{2\perp}'k_\perp^2}
{\left[\beta k_\perp^2+(1-\beta)k_{1\perp}^{\prime 2}\right]
\left[\beta k_\perp^2+(1-\beta)k_{2\perp}^{\prime 2}\right]}\ , \label{eq:tq}
\end{eqnarray}
with  $k_{i\perp}'=k_\perp-k_{i\perp}$. $\mathcal{F}_{x_{I\!\!P}}(k_{i\perp}, \Delta_\perp)$ is the Fourier transform of the dipole S-matrix in the fundamental representation,
\begin{eqnarray}
&&\mathcal{F}_x(q_\perp, \Delta_\perp)=\int\frac{d^2b_\perp d^2r_\perp}{(2\pi)^4} e^{iq_\perp \cdot r_\perp +i \Delta_\perp\cdot b_\perp}\nonumber\\
&&\times \frac{1}{N_{c}}\left\langle\text{Tr}\left[
U\left( b_{\perp }+\frac{r_\perp}{2}\right) U^{\dagger }\left( b_{\perp }-\frac{r_\perp}{2} \right)\right]\right\rangle_x \label{g2} \ , 
\end{eqnarray}
where 
\beq
U(b_\perp)={\rm P}\exp\left(ig\int_{-\infty}^\infty dz^-A^+(z^-,b_\perp)\right), \label{wilson}
\eeq
is the Wilson line along the lightcone in the fundamental representation. 
Similarly, the diffractive gluon DPDF is represented by the right diagram in Fig.~\ref{fig:dff} and reads 
\begin{eqnarray}
&&x\frac{d f_g^D(\beta,k_\perp;x_{I\!\!P})}{dY_{I\!\!P} dt} =  \int
d^2k_{1\perp}d^2k_{2\perp} {\cal G}_{x_{I\!\!P}}(k_{1\perp},\Delta_\perp)
\nonumber \\ 
&&~~\times{\cal G}_{x_{I\!\!P}}(k_{2\perp},\Delta_\perp)\frac{N_c^2-1}{\pi(1-\beta)}  {\cal T}_g(k_\perp,k_{1\perp},k_{2\perp})\ ,\label{eq:dffgluon}
\end{eqnarray}
where we again defined ${\cal T}_g\equiv 
T_g(k_\perp,k_{1\perp},k_{2\perp})-T_g(k_\perp,0,k_{2\perp})-T_g(k_\perp,k_{1\perp},0)+T_g(k_\perp,0,0)$ with 
\begin{eqnarray}
&&    T_g(k_\perp,k_{1\perp},k_{2\perp})=\frac{1}{\left[\beta k_\perp^2+(1-\beta)k_{1\perp}^{\prime 2}\right]}\nonumber\\
&&~~~\times \frac{1}{
\left[\beta k_\perp^2+(1-\beta)k_{2\perp}^{\prime 2}\right]}\left[\beta(1-\beta){k_\perp^2}\frac{k_{1\perp}^{\prime 2}+k_{2\perp}^{\prime 2}}{2}\right.\nonumber\\
&& ~~~\left.+(1-\beta)^2 (k_{1\perp}'\cdot k_{2\perp}')^2+\beta^2 \frac{(k_\perp^2)^2}{2}\right]
\ . \label{eq:tg}
\end{eqnarray}
The {\it gluon} dipole S-matrix is defined as 
\begin{eqnarray}
&&\mathcal{G}_x(q_\perp, \Delta_\perp)=\int\frac{d^2b_\perp d^2r_\perp}{(2\pi)^4} e^{iq_\perp \cdot r_\perp +i \Delta_\perp\cdot b_\perp}\nonumber\\
&&\times \frac{1}{N^2_{c}-1}\left\langle\text{Tr}\left[
\widetilde{U}\left( b_{\perp }+\frac{r_\perp}{2}\right) \widetilde{U}^{\dagger }\left( b_{\perp }-\frac{r_\perp}{2} \right)\right]\right\rangle_x \  ,\label{g2g}
\end{eqnarray}
where $\widetilde{U}$ is the same Wilson line but in the adjoint representation. 

The above DPDFs can be applied to  semi-inclusive diffractive processes (Fig.~\ref{fig:siddis}) where the argument for QCD  factorization  should be analogous to those for hard diffractive DIS~\cite{Collins:1997sr} and  non-diffractive semi-inclusive DIS~\cite{Collins:2011zzd,Ji:2004wu}. The combination of these two factorization proofs should lay the groundwork  for the QCD factorization of our process.  
As an example, consider semi-inclusive quark production. To leading order, the differential cross section can be immediately written down in terms of the quark DPDF 
\begin{equation}
    \frac{d\sigma^{\rm SIDDIS}(\ell p\to \ell'p' qX)}{dx_Bdyd^2k_\perp dY_{I\!\!P}dt}=\sigma_0 e_q^2x_B\frac{df_q^D(\beta,k_\perp;x_{I\!\!P})}{dY_{I\!\!P}dt}\,, \label{eq:siddis}
\end{equation}
where $\sigma_0=\frac{4\pi\alpha_{em}^2S_{ep}}{Q^4}\left(1-y+\frac{y^2}{2}\right)$ and the usual DIS variables are   defined as $x_B=Q^2/2P\cdot q$, $y=q\cdot P/k_\ell\cdot P$ and $S_{ep}=(k_\ell+P)^2$.  Additional soft factors  will be needed at higher orders. The formula (\ref{eq:siddis}) is consistent with the direct calculation within the CGC formalism where the cross section $\gamma^*p \to qXp'$ is obtained by first considering the split $\gamma^* \to q\bar{q}$ and then  integrating over the phase space of the antiquark. This is demonstrated in Appendix B using a technique  developed in \cite{Marquet:2009ca}. We expect that the consistency persists to higher orders starting the $q\bar q g$ final state, but this has to be carefully investigated in future work.

The quark and gluon dipole S-matrices may contain 
nontrivial correlations between $\Delta_\perp$ and $k_\perp$ with observable consequences. 
Especially if these correlations depend on the nucleon spin, they will open up new opportunities to explore  spin-orbital correlations inside hadrons, 
\begin{itemize}
     \item The $\cos(2\phi)$ correlation \cite{Hatta:2016dxp} between $\Delta_\perp$ and $k_{i\perp}$ in the dipole S-matrix results in a similar correlation between $k_\perp$ and $\Delta_\perp$ in the DPDFs. This can be observed experimentally as $\cos 2\phi$  and higher order azimuthal asymmetries between the recoiling proton and observed hadrons in SIDDIS in Fig.~\ref{fig:siddis}. Previously, such asymmetries  have been studied theoretically in  exclusive processes~\cite{Hatta:2016dxp,Mantysaari:2019hkq}.

     \item For non-diffractive processes, the leading order TMDs contain correlations between the transverse momentum and the polarizations of the parton and/or the nucleon states~\cite{Mulders:1995dh,Boer:1997nt,Bacchetta:2006tn}. Extending these parameterizations to the DPDFs will provide a unique perspective of the hadron tomography.
    For example, 
    it has been shown that the TMD quark and gluon Sivers functions at small-$x$ are related to the spin-dependent odderon~\cite{Zhou:2013gsa,Boer:2015pni,Boussarie:2019vmk,Dong:2018wsp,Kovchegov:2021iyc}. It is interesting to explore how the diffractive Sivers functions 
    arise from the spin-dependent dipole S-matrix. 
    \end{itemize}

In addition, the gluon GPDs can be expressed in terms of the dipole amplitudes~\cite{Hatta:2017cte}:
$H_g(x_{I\!\!P},\Delta_\perp) =\frac{2N_c}{\alpha_s}  \int d^2 q_\perp q_\perp^2 {\cal F}_{x_{I\!\!P}}(q_\perp,\Delta_\perp)=\frac{N_c^2-1}{N_c\alpha_s}  \int d^2 q_\perp q_\perp^2 {\cal G}_{x_{I\!\!P}}(q_\perp,\Delta_\perp)$, where we have set the skewness parameter in the GPD $\xi=x_{I\!\!P}$. 
Therefore, we can rewrite the DPDFs in terms of the gluon GPD as well. In particular, for large transverse momentum DPDFs, we have
\begin{equation}
\left.x\frac{d f_{q,g}^D(\beta,k_\perp;x_{I\!\!P})}{dY_{I\!\!P} dt}\right|_{k_\perp\gg Q_s} = \frac{\alpha_s^2}{2\pi}\frac{{\cal C}_{q,g}}{k_\perp^4}\left(H_g(x_{I\!\!P},\Delta_\perp)\right)^2 \ ,
\end{equation}
where ${\cal C}_q= \beta^3(1-\beta)^2/N_c$ and ${\cal C}_g=(1+2\beta)^2(1-\beta)^3 N_c^2/(N_c^2-1)$. This builds an interesting connection between hard diffractive processes and the GPD physics. In addition, the $1/k_\perp^4$ power behavior is very different from that of the non-diffractive quark and gluon  TMDs~\cite{Marquet:2009ca,Xiao:2017yya} which behave as $1/k_\perp^2$ at large-$k_\perp$ leading to logarithmically divergent $k_\perp$ integrals.

\section{Unpolarized DPDFs in a saturation model}

\begin{figure}[t]
\begin{center}
\includegraphics[width=0.35\textwidth]{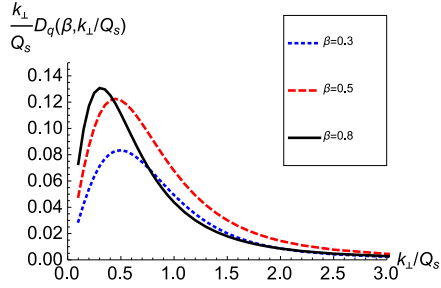}
\end{center}
\caption[*]{The transverse momentum dependence of the quark diffractive distribution for different values of $\beta=0.3,0.5,0.8$, respectively, plotted as functions of $k_\perp/Q_s$, see, Eq.~(\ref{eq:e6}). }
\label{fig:dffquark0}
\end{figure}

Here, we consider  the kinematics of zero momentum transfer from the target ($\Delta_\perp=0)$ and investigate  the $\beta$-dependence of   DPDFs $f^D_{q,g}$ in detail.  To illustrate that, we evaluate the $k_{i\perp}$ integrals in (\ref{dffquark}) and (\ref{eq:dffgluon}) numerically assuming a simple Gaussian form for the color dipole  (see, e.g., \cite{Golec-Biernat:1999qor}) 
\begin{equation}
    {\cal F}_{x_{I\!\!P}}(k_{i\perp})=\frac{S_\perp}{(2\pi)^2}\frac{1}{\pi Q_s^2}e^{-k_{i\perp}^2/Q_s^2}\ ,
\end{equation}
where the (quark) saturation scale $Q_s$ depends on $x_{I\!\!P}$ and $S_\perp$ represents the transverse area of the target. The same parameterization will be used for the gluon dipole ${\cal G}_{x_{I\!\!P}}(k_{i\perp})$ but with the saturation momentum for the adjoint representation $Q_{as}$. They are related as  $Q_{as}^2=\frac{C_A}{C_F} Q_s^2\approx 2 Q_s^2$. With these assumptions, we find that the DPDFs depend on the ratios  $k_\perp/Q_s$ and $k_\perp/Q_{as}$ for the quark and gluon distributions, respectively,
\begin{eqnarray}
x\frac{d f_{q,g}^D(\beta,k_\perp;x_{I\!\!P})}{dY_{I\!\!P} dt} = {\cal N}_{q,g} D_{q,g}\left(\beta,\frac{k_\perp}{Q_{s,as}}\right) \ ,\label{eq:e6}
\end{eqnarray}
where ${\cal N}_q=S_\perp^2N_c/(2\pi)^5$ and ${\cal N}_g=S_\perp^22(N_c^2-1)/(2\pi)^5$. For ordinary TMDs, relations like (\ref{eq:e6}) exhibit the phenomenon of geometric scaling, namely, distributions $f(x,k_\perp)$ depend on $k_\perp$ and $x$ only through the ratio $k_\perp/Q_s(x)$. However, in the present problem, the extra factor $\beta$ complicates this interpretation.   
In Fig.~\ref{fig:dffquark0}, we show  the quark DPDF $D_q(\beta,k_\perp/Q_s)$ as functions of $k_\perp/Q_s$ for different values of $\beta$. The strong falloff at large $k_\perp$ confirms the above power counting analysis. On the other hand,  the shape and magnitude of these curves strongly depend on $\beta$.  

\begin{figure}[t]
\begin{center}
\includegraphics[width=0.35\textwidth]{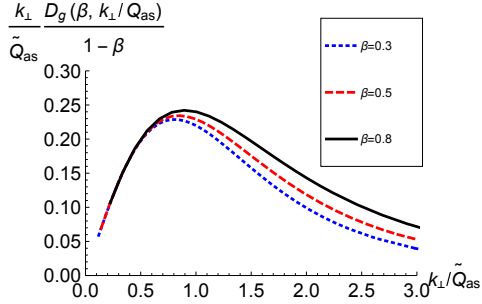}
\end{center}
\caption[*]{Same as Fig.~\ref{fig:dffquark0} for the gluon. Here, we plot $\frac{k_\perp}{\widetilde{Q}_{as}}\frac{ D(\beta,k_\perp/Q_{as})}{1-\beta}$ as functions of $k_\perp/\widetilde Q_{as}$ for $\beta=0.3,0.5,0.8$ (blue, red, black curves), respectively, where $\widetilde{Q}_{as}^2=(1-\beta)Q_{as}^2$.}
\label{fig:dffgluon}
\end{figure}

Nonetheless, the authors of Ref.~\cite{Iancu:2021rup} have observed that the DPDFs do exhibit geometric scaling if it is expressed in terms of the modified saturation momentum   $\widetilde Q_{(a)s}^2\equiv (1-\beta)Q_{(a)s}^2$. A simple explanation within our approach is to look at  the denominator of 
(\ref{eq:tq}) and (\ref{eq:tg}) 
\beq
\frac{1}{ k_\perp^2+(1-\beta)k_{i\perp}^2-2(1-\beta)k_{i\perp}\cdot k_\perp}\,,
\label{life}
\eeq
and noting that, typically, $k_{i\perp}\sim Q_{(a)s}$. When $1-\beta$ is order unity, the characteristic value of $k_\perp$ is  $k_{i\perp}\sim Q_{(a)s}$ such that the scaling geometric variable is $k_\perp/Q_{(a)s}$. However, when $1-\beta$ becomes very small, (\ref{life}) is power suppressed when 
$k_\perp^2\gtrsim (1-\beta)k^2_{i\perp}\sim (1-\beta)Q_{as}^2=\widetilde{Q}^2_{(a)s}$. Thus the $k_\perp$-distribution is effectively limited to $k_\perp\lesssim \widetilde{Q}_{(a)s}$ and the scaling variable becomes $k_\perp/\widetilde{Q}_{(a)s}$. 

To corroborate the above argument, in Fig.~\ref{fig:dffgluon} we plot $D_g$ as a function of $k_\perp/\widetilde{Q}_{as}$ for the same three values of $\beta$ as in Fig.~\ref{fig:dffquark0}. We divided the results by the common prefactor $1-\beta$, which naturally arises from the large-$\beta$ analysis (see below). The three curves now agree very well with each other and peak around the modified saturation momentum $k_\perp =\widetilde{Q}_{as}$, in agreement with   \cite{Iancu:2021rup}.

\begin{figure}[t]
\begin{center}
\includegraphics[width=0.3\textwidth]{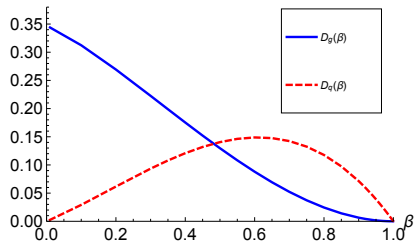}
\end{center}
\caption[*]{The integrated quark and gluon diffractive distribution functions as functions of $\beta$. Relative sizes between these two depend on additional factors, see, Eq.~(\ref{eq:intdff}).}
\label{fig:dffquarkgluon}
\end{figure}

We further integrate out $k_\perp$ to derive the integrated DPDFs and compare to previous results~\cite{Iancu:2021rup,Golec-Biernat:2001gyl}. Within the Gaussian approximation for the dipole amplitudes, we can write  
\begin{eqnarray}
x\frac{d\, f_{q,g}^D(\beta;x_{I\!\!P})}{dY_{I\!\!P} dt} = {\cal N}_{q,g} 2\pi {\cal D}_{q,g}(\beta)Q_{s,as}^2 \ .\label{eq:intdff}
\end{eqnarray}
In Fig.~\ref{fig:dffquarkgluon}, we show the numerical results for ${\cal D}_{q,g}(\beta)$. In order to understand the different behaviors in the quark and gluon cases, we will derive below some analytic results around the endpoint regions $\beta=0$ and $\beta=1$. This also helps to provide simple approximate expressions for the functions ${\cal D}_{q,g}(\beta)$ in the whole kinematic interval $0\le \beta \le 1$,
\begin{eqnarray}
    {\cal D}_q(\beta)&=&\left(b_1+b_2(1-\beta)\right) \beta(1-\beta)\ , \label{betaq}\\
{\cal D}_g(\beta)&=&(a_0+a_1\beta)(1-\beta)^2\ , \label{beta2}
\end{eqnarray}
with $b_1=\frac{3\pi^2}{16}-1$, $b_2=\frac{20-3\pi^2}{16}$, $a_0=\frac{\ln(2)}{2}$ and $a_1=\frac{45\pi^2-272}{256}-\frac{\ln(2)}{2}$. These parameters are determined by the endpoint behaviors.
When $\beta=0$, by averaging over the azimuthal angles of $k_{1\perp}$ and $k_{2\perp}$ in Eq.~(\ref{dffquark},\ref{eq:dffgluon}), we find
\begin{eqnarray}
&&{\cal T}_q|_{\beta\to 0}=\Theta(k_{1\perp}-k_\perp)\times \Theta(k_{2\perp}-k_\perp)\ ,\\
&&{\cal T}_g|_{\beta\to 0}=\frac{1}{8k_\perp^4}\left(k_\perp^2+k_{1\perp}^2+(k_\perp^2-{k}_{1\perp}^2){\rm Sign}({k}_{1\perp}-k_\perp)\right)\nonumber\\
&&~~\times \left(k_\perp^2+ k_{2\perp}^2+(k_\perp^2-{k}_{2\perp}^2){\rm Sign}({k}_{2\perp}-k_\perp)\right) \ .
\end{eqnarray} 
The subsequent integrals can be done analytically and this fixes the values of $b_1+b_2$ and $a_0$. 

On the other hand, the behavior near $\beta\to 1$ is much more complicated. The integrand of Eq.~(\ref{dffquark}) vanishes if we set $\beta=1$ naively. In order to obtain the correct leading behavior in $1-\beta$, we first make the rescaling  $\tilde k_{i\perp}=\sqrt{1-\beta}k_{i\perp}$ after which $k_\perp$ and $\tilde{k}_{i\perp}$ become comparable (see the argument around (\ref{life})).  We  then expand the integrand ${\cal T}_{q,g}(k,k_{1\perp},k_{2\perp})$ around $\beta=1$,
\begin{eqnarray}
   {\cal T}_q|_{\beta\to 1}&=&\frac{\tilde k_{1\perp}^2 (2k_\perp^2 + \tilde k_{1\perp}^2)}{(k_\perp^2+\tilde k_{1\perp}^2)^2}\frac{\tilde k_{2\perp}^2 (2k_\perp^2 + \tilde k_{2\perp}^2)}{(k_\perp^2+\tilde k_{2\perp}^2)^2}\ ,\\
   {\cal T}_g|_{\beta\to 1}&=&\frac{(1-\beta)^2}{2} \frac{\tilde{k}_{1\perp}^2(3k_\perp^4+3k_\perp^2\tilde k_{1\perp}^2+\tilde{k}_{1\perp}^4)}{(k_\perp^2+\tilde k_{1\perp}^2)^3}\nonumber\\
&&\times \frac{\tilde{k}_{2\perp}^2(3k_\perp^4+3k_\perp^2\tilde k_{2\perp}^2+\tilde{k}_{2\perp}^4)}{(k_\perp^2+\tilde k_{2\perp}^2)^3} \ .
\end{eqnarray}
The remaining integrals over $\widetilde{k}_{i\perp}$ and $k_\perp$ can be performed analytically. An overall factor of $1-\beta$ comes from the final integral $\int d^2k_\perp \sim \widetilde{Q}_{s}^2 =(1-\beta)Q_s^2$, and the parameters $b_1$ and $a_0+a_1$ can be read off from the coefficients.  

Similar parameterizations for the quark/gluon DPDFs have been derived in Ref.~\cite{Buchmuller:1998jv}. Our result in Eq.~(\ref{betaq}) for the quark case agrees with theirs, whereas there is a factor of 2 difference for the gluon case in Eq.~(\ref{beta2}). As we explain in detail in Appendix \ref{appdpdf}, we also find agreement with the results of Refs.~\cite{Golec-Biernat:1999qor,Golec-Biernat:2001gyl}, although in the gluon case there is an overall factor 2 difference due to an inconsistent parameterization of the dipole S-matrix in the adjoint representation in~\cite{Golec-Biernat:1999qor}, see also the discussion in~\cite{Marquet:2007nf}. This also explains the difference between our result and that from Ref.~\cite{Buchmuller:1998jv} for the gluon DPDF.

The above results can be regarded as inputs for the DPDFs at the initial scale $\mu\sim Q_{s,as}$. They have also been extracted from the HERA experiments~\cite{H1:1997bdi,ZEUS:1997fox,Newman:2013ada}. The shapes and magnitudes of the quark/gluon diffractive PDFs calculated from the CGC/dipole formalism in Eqs.~(\ref{betaq},\ref{beta2}) are very similar to those determined from these measurements. Of course, to have a quantitative comparison, one has to evolve our results to the relevant scales and compare them to those in Refs.~\cite{H1:1997bdi,ZEUS:1997fox,Newman:2013ada}. We will pursue this direction in a future publication. 

In addition, we notice that the ratio between total momentum fractions carried by quarks and gluons, $\frac{2N_f\int d\beta xf^D_q(\beta;x_{I\!\!P})}{\int d\beta xf^D_g(\beta;x_{I\!\!P})}\approx 0.33$ with $N_f=3$, is very close to that  determined by the ZEUS Collaboration~\cite{ZEUS:2009uxs}. The additional difference can be explained by the evolution effect from $Q_s$ to the scales used there. It is interesting to further notice that the total momentum fraction is normalized,
\begin{equation}
    \int d\beta x
    \left[\frac{2N_fdf^D_q(\beta;x_{I\!\!P})}{dY_{I\!\!P}dt}+\frac{df^D_g(\beta;x_{I\!\!P})}{dY_{I\!\!P}dt}\right]\approx \frac{S_\perp^2}{(2\pi)^3}Q_s^2\ ,
\end{equation}
where we have applied $Q_{as}^2/Q_s^2\approx C_A/C_F$. 

\section{Conclusion}

In summary, we have established a connection between the QCD factorization for hard diffractive processes and the small-$x$ CGC formalism by computing the diffractive PDFs in terms of the dipole S-matrices. The transverse momentum dependent DPDFs can be studied in  semi-inclusive diffractive DIS processes. The integrated DPDFs have been evaluated with a Gaussian assumption for the dipole S-matrix. These results can be regarded as inputs at the initial scale  for collinear QCD factorizations applied  to all other hard diffractive processes. This provides a powerful tool to unify different frameworks at small-$x$.

We have briefly commented on the nontrivial correlation between the momentum transfer $\Delta_\perp$ and the parton transverse momentum $k_\perp$ in the DPDFs as a direct consequence of the correlations in the dipole amplitude between $\Delta_\perp$ and $k_{i\perp}$ \cite{Hatta:2016dxp}. Such correlations are measurable at the EIC where the proton recoil momentum $\Delta_\perp$ can be directly measured by  Roman pots. Moreover, if these correlations depend on the nucleon spin, they will open up new opportunities to explore the spin-orbital correlations inside the nucleon. This may lead to a complementary method to investigate the gluon orbital angular momentum contribution to the proton spin, as compared to the proposals of exclusive processes in the literature for this purpose~\cite{Ji:2016jgn,Hatta:2016aoc,Bhattacharya:2022vvo,Courtoy:2013oaa,Courtoy:2014bea}. 
We will come back to these questions in a future publication.

{\bf Acknowledgments:} We thank Edmond Iancu and Al Mueller for the discussions. This material is based upon work supported by the LDRD programs of LBNL and Brookhaven Science Associates, and by the U.S. Department of Energy, Office of Science, Office of Nuclear Physics, under contract numbers DE-AC02-05CH11231 and DE- SC0012704. This work is also supported by the National Natural Science Foundations of China under Grant No. 11575070.

\appendix
\begin{widetext}

\section{ Calculation of DPDFs}

In this appendix we give the outline of the derivation of the quark DPDF (\ref{dffquark}) and gluon DPDF (\ref{eq:dffgluon}). To compute the gluon DPDF from its definition of (\ref{tmdung}), we first expand the gluonic operator $F^{+\mu}$ and the associated gauge links ${\cal L}_n$ which interact with the background field $A^+$ of the  CGC.  The multiple interactions with the CGC can be resummed by the  Wilson lines $U$ which constitute the dipole S-matrix. There are important differences compared to the non-diffractive TMD gluon distributions computed in Ref.~\cite{Xiao:2017yya}. First, there is no leading order correspondence between the gluon DPDF and the dipole amplitude as was the case for the non-diffractive TMD gluon distribution~\cite{Dominguez:2010xd,Dominguez:2011wm}. This is because a final state gluon radiation is required to guarantee the colorless exchange between the gluonic fields and the nucleon target in the diffractive case. As a consequence,  non-zero contributions come from the diagrams shown in Fig.~\ref{fig:dff}. Second, because of the colorless exchange, the amplitude of this diagram depends on the dipole S-matrix, whereas for the non-diffractive TMDs, it is the amplitude squared that depends on the dipole S-matrix.

Now, let us move to the derivation. We closely follow the calculation of the non-diffractive TMD gluon distribution at small-$x$ in Ref.~\cite{Xiao:2017yya}. The major difference, as mentioned above, is the colorlesss exchange between the nucleon target the partonic part from the gluon distribution calculations, see, Fig.~\ref{fig:dff}. Consider the right diagram in Fig.~\ref{fig:dff}. We first observe that the gluon connecting the target and the Wilson line (denoted by a double line) has a vanishing plus-momentum. This is a consequence of the eikonal approximation (note the $dz^-$ integral in (\ref{wilson})). Therefore, the other gluon entering the triple gluon vertex has a plus-momentum  $k_1^+=x_{I\!\!P}P^+$. The kinematics of this  splitting is 
\beq
k_1^\mu=\left(k_1^+,0,k_{1\perp }\right) \to k^\mu+q^\mu= (\beta k_1^+,-q^-,k_{1\perp}-k_\perp) + \left((1-\beta)k_1^+,q^-,k_\perp\right).
\eeq
Since the outgoing gluon with momentum $q^\mu$ is on-shell and transverse, 
\beq
q^-=\frac{k_\perp^2}{2(1-\beta)k_1^+},\qquad \epsilon^* \cdot q = 0, \quad \to \quad  \epsilon^{+*} =\frac{\vec{\epsilon}^*_\perp \cdot \vec{k}_\perp}{q^-},
\eeq
where $\epsilon^\alpha$ is the polarization vector, and we work in the gauge $\epsilon^-=0$. The phase space integral for the emitted gluon is thus  
\beq
\frac{d^4q}{(2\pi)^3} \delta(q^2)= 
\frac{d^2k_\perp}{16\pi^3}\frac{dx_{I\!\!P}}{(1-\beta)x_{I\!\!P}} = \frac{d^2k_\perp}{16\pi^3}\frac{dY_{I\!\!P}}{1-\beta}.
\eeq
The intermediate gluon denominator takes the form 
\beq
\frac{1}{k^2}=\frac{1}{-2\beta k_1^+q^- - (\vec{k}_{1\perp}-\vec{k}_\perp)^2} = -\frac{1-\beta}{\beta\vec{k}_{\perp}^2 +(1-\beta)(\vec{k}_{1\perp}-\vec{k}_\perp)^2}. \label{propglue}
\eeq
Triple gluon vertex can be evaluated as  
\beq
\int d^3z e^{ik\cdot z}\langle q,\alpha|F^{+\nu}(z) &\sim& \langle q,\alpha|\left(\int d^4x A AA \right)(k^+A^\nu-k^\nu A^+)  \nn
&\sim& \biggl[k^+(g^{\nu\alpha}_\perp(-k+q)^- + g^{\alpha-}(-2q-k)^\nu+g^{-\nu}(2k+q)^\alpha) \nn 
&&  \qquad -k^\nu (g^{+\alpha}(-k+q)^- + g^{\alpha-}(-2q-k)^++g^{-+}(2k+q)^\alpha) \biggr]\epsilon^*_\alpha(q)A^+(k_g) \nn
&=& \left(2k^+q^- g^{\nu\alpha}_\perp\epsilon^*_\alpha -2q^-k^\nu \epsilon^{+*} - 2k^\nu k\cdot \epsilon^* \right)A^+ \nn
&=&-\frac{1}{1-\beta}\left( \beta k_{\perp}^2\delta^{\nu\alpha}_\perp +2(1-\beta)(k_{1\perp}^\nu-k^\nu_\perp) (k_{1\perp}^\alpha-k^\alpha_{\perp})\right) \epsilon^*_{\alpha}A^+, \label{triple}
\eeq
where $\nu$ is transverse, and we used 
\beq
k\cdot \epsilon^* = -q^-\epsilon^{+*} -(k_{1\perp}-k_\perp) \cdot \vec{\epsilon}^*_\perp =-k_{1\perp}\cdot \epsilon^*_\perp .
\eeq
Multiplying (\ref{propglue}) by (\ref{triple}) and squaring, we obtain the kernel in (\ref{eq:tg}). 

Similarly, we can derive the TMD quark DPDF shown in left panel of Fig.~\ref{fig:dff}. In particular, the partonic part is the same as that in Eq.~(41) of the published version of Ref.~\cite{Xiao:2017yya}. Again, to compute the quark DPDF, we need to make sure that the colorless exchange between the partonic part and the nucleon target. This requires the following replacement,
\begin{eqnarray}
    \frac{1}{N_c} {\rm Tr}\left[\langle U(x_1)U^\dagger(x_2)U(y_2)U^\dagger(y_1)\rangle- \langle
U(x_1)U^\dagger(x_2)\rangle-\langle U(x_2)U^\dagger(y_1)\rangle+1\right]\nonumber\\
\longrightarrow  \frac{1}{N_c} {\rm Tr}\left[\langle U(x_1)U^\dagger(x_2)\rangle-1\right]\times \frac{1}{N_c}{\rm Tr}\left[ \langle U(y_2)U^\dagger(y_1)\rangle-1\right]\ .
\end{eqnarray}
With additional kinematic variable replacements, we  arrive at the result in Eq.~(\ref{dffquark}). 
    
\section{Derivation of Eq.~(\ref{eq:siddis})}

In this appendix, we  show how the semi-inclusive quark production in diffractive DIS calculated in the CGC/dipole formalism can be factorized into the TMD quark DPDF in the limit of $k_\perp\ll Q$, where $k_\perp$ is final state quark transverse momentum and $Q$ the virtuality of the photon. In this process, $\ell+p\to \ell'+k+X+p'$, the incoming lepton radiates a highly virtual photon, which interacts with the nucleon target diffractively and produces a final state quark with momentum $k$. The derivation here also applies to semi-inclusive hadron production in diffractive DIS, where a final state fragmentation function will be included. At this order, there is no difference between  hadron and jet productions. At higher orders, the TMD fragmentation and jet functions will enter the factorization formula, respectively.  

In the CGC/dipole formalism, the quark production comes from the process that the virtual photon splits into a quark-antiquark pair and goes through diffractive interaction with the nucleon target, $\gamma^*p\to q\bar q p$, at the leading order. The amplitude for this process has been computed~\cite{Hatta:2016dxp}, and the differential cross section for the quark production can be derived by integrating out the phase space of the antiquark,
\begin{eqnarray}
\frac{d\sigma^{\rm SIDDIS}(\ell p\to \ell'qp'X)}{dx_BdQ^2d^2k_\perp dY_{I\!\!P}dt}
&=&\frac{\alpha_{em}^2e_q^2N_c}{x_BQ^2}\left(1-y+\frac{y^2}{2}\right)\int{dz}\delta(1-\beta-\beta\frac{k_\perp^2}{\epsilon_f^2})\int d^2 q_\perp d^2 q_\perp^\prime  \mathcal{F}_x(q_\perp, \Delta_\perp)\mathcal{F}_x(q_\perp^\prime , \Delta_\perp) \notag \\
&&\times \left(z^2+(1-z)^2\right) \left[  \frac{k_\perp}{k_\perp^2 +\epsilon_f^2 } - \frac{k_\perp-q_\perp }{(k_\perp-q_\perp)^2 +\epsilon_f^2} \right] \cdot  \left[  \frac{k_\perp}{k_\perp^2 +\epsilon_f^2 }  - \frac{k_\perp-q_\perp^\prime}{(k_\perp-q_\perp^\prime)^2 +\epsilon_f^2 } \right]   \ ,\label{eq:dct} 
\end{eqnarray}
where $\beta=x_B/x_{I\!\!P}$ and $z$ is the momentum fraction of incoming photon carried by the final state quark and $\epsilon_f^2=z(1-z)Q^2$. The Delta function in the above equation comes from the momentum conservation along the nucleon momentum direction. To derive that, we notice that the momentum fractions of the incoming nucleon carried by the quark and antiquark: $x_q=\frac{x_Bk_\perp^2}{zQ^2}$ and $x_{\bar q}=\frac{x_Bk_\perp^2}{(1-z)Q^2}$, respectively, where we have applied the approximation of $\Delta_\perp\approx 0$ and the quark and antiquark have balanced transverse momenta $|k_{q\perp}|\sim |k_{\bar q\perp}|$. The momentum conservation leads to $x_{I\!\!P}=x_B+x_q+x_{\bar q} $, which results in ${(1-\beta)}/{\beta}={k_\perp^2}/{\epsilon_f^2}$.

Similar to the non-diffractive quark production in DIS process~\cite{Marquet:2009ca}, the above contribution is also dominated by the so-called aligned jet configuration, i.e., $z\sim 1$ or $z\sim 0$. Only in this kinematics can we find that $\epsilon_f^2\sim k_\perp^2$ and the differential cross section does not vanish at large $Q^2$. This can be illustrated by rewriting the above Delta function,
\begin{eqnarray}
    \delta\left(1-\beta-\beta \frac{k_\perp^2}{\epsilon_f^2}\right)&=&\frac{\beta}{(1-\beta)^2}\frac{k_\perp^2}{Q^2}\delta\left(z(1-z)-\frac{\beta}{1-\beta}\frac{k_\perp^2}{Q^2}\right) \nonumber\\
    &=&\frac{\beta}{(1-\beta)^2}\frac{k_\perp^2}{Q^2}\left(\frac{\delta(1-z)}{z}+\frac{\delta(z)}{1-z}\right)\ ,
\end{eqnarray}
where the last equation comes from small $k_\perp^2/Q^2$ expansion. Substituting the above expansion result into Eq.~(\ref{eq:dct}), we derive Eq.~(\ref{eq:siddis}).

\section{Comparison to the diffractive structure functions calculated in Ref.~\cite{Golec-Biernat:1999qor}}
\label{appdpdf}

Integrating over the azimuthal angles in Eqs.~(\ref{dffquark}) and (\ref{eq:dffgluon}), we find 

\begin{eqnarray}
x\frac{d f_q^D(\beta,k_\perp;x_{I\!\!P})}{dY_{I\!\!P}dt} &=& \frac{\pi N_c \beta}{8(1-\beta)^2} \left[ \int_0^\infty dk_{1\perp}^2{\cal F}(k_{1\perp}^2) \left\{ 1-2\beta + \frac{(1-\beta)k_{1\perp}^2 -(1-2\beta)k_\perp^2}{\sqrt{(k_\perp^2+(1-\beta)k_{1\perp}^2)^2-4(1-\beta)^2k_\perp^2 k_{1\perp}^2}}\right\}\right]^2\ , \label{angleint1} \\
x\frac{d f_g^D(\beta,k_\perp;x_{I\!\!P})}{dY_{I\!\!P}dt} &=& \frac{\pi (N_c^2-1)}{8(1-\beta)^3} \Biggl[ \int_0^\infty dk_{1\perp}^2{\cal G}(k_{1\perp}^2) \Biggl\{ \beta^2+(1-\beta)^2 + \frac{k_{1\perp}^2}{k_\perp^2}(1-\beta) \nn && \qquad \qquad \qquad -\frac{((1-2\beta)k_\perp^2-(1-\beta)k_{1\perp}^2)^2+2\beta(1-\beta)k_\perp^4 }{k_\perp^2\sqrt{(k_\perp^2+(1-\beta)k_{1\perp}^2)^2-4(1-\beta)^2k_\perp^2 k_{1\perp}^2}}\Biggr\}\Biggr]^2\ . \label{angleint2}
\end{eqnarray}
We immediately recognize the same structure as in  the diffractive structure functions calculated in Ref.~\cite{Golec-Biernat:1999qor}. For the $q\bar{q}$ contribution to the diffractive structure function for the transversely polarized photon, we can explicitly rewrite Eq.~(20) of Ref.~\cite{Golec-Biernat:1999qor} as
\beq
F^D_{\{t,q\bar{q}\}}(Q^2,\beta,x_{I\!\!P}) = Q^2\pi (1-\beta) \int_0^1 d\alpha (\alpha^2+(1-\alpha)^2)  \frac{df_q^D(\beta,k_\perp;x_{I\!\!P})}{dY_{I\!\!P}} \label{gb}
\eeq
after adjusting the normalization difference of the unintegrated gluon distribution 
\begin{eqnarray}
    {\cal F}_{{\rm Ref.}[26]
    }(q_\perp)&=&\frac{2\pi N_cq_\perp^2}{\alpha_s}{\cal F}(q_\perp) \ .
\label{ktfact}
\end{eqnarray}
 In (\ref{gb}), we have integrated over $t$ assuming the exponential form $e^{B_D t}$ (as was done in \cite{Golec-Biernat:1999qor}). The parameter $\alpha$ is related to $k_\perp$ as 
\begin{equation}
    k_\perp^2=\alpha(1-\alpha)Q^2\frac{1-\beta}{\beta} \ .
\end{equation}
Eq.~(\ref{gb}) can then be recognized as a $k_T$ factorization formula. At large-$Q^2$, one can take the collinear limit of this by inserting the above constraint and expanding in $k_\perp^2/Q^2$ 
\begin{eqnarray}
1&=&    \int dk_\perp^2\delta\left(k_\perp^2-\alpha(1-\alpha)\frac{Q^2(1-\beta)}{\beta}\right) 
\approx \frac{\beta}{Q^2(1-\beta)}\left(\frac{\delta(\alpha)}{1-\alpha}+\frac{\delta(1-\alpha)}{\alpha}\right)\int dk_\perp^2\ .
\end{eqnarray}
Now the integral over $\alpha$ can be easily carried out, giving
\beq
F^D_{\{t,q\bar{q}\}}(Q^2,\beta,x_{I\!\!P}) \approx  2\beta\frac{df_q^D(\beta;x_{I\!\!P})}{dY_{I\!\!P}}\,,
\eeq
where the factor of 2 accounts for the quark and antiquark contributions.  The longitudinal diffractive structure function $F^D_{l,q\bar{q}}$, like the inclusive longitudinal structure function, is power suppressed. Therefore,  Eq.~(21) of Ref.~\cite{Golec-Biernat:1999qor}  does not have a corresponding interpretation in terms of the DPDF. 

On the other hand, the $q\bar{q} g$ contribution to the transverse structure function, Eq.~(23) of \cite{Golec-Biernat:1999qor}, is related to the gluon DPDF (\ref{angleint2}). A direct comparison is somewhat obscure because   Ref.~\cite{Golec-Biernat:1999qor} did not distinguish the quark and gluon dipole amplitudes (and hence the corresponding saturation momenta $Q_s$ and $Q_{as}$). Yet, we can make the following identification in the gluon case 
\begin{eqnarray}
    {\cal F}_{{\rm Ref.}[26]
    }(q_\perp)\to \frac{\pi (N_c^2-1)q_\perp^2}{N_c\alpha_s}{\cal G}(q_\perp) \ .
    \label{glueid}
\end{eqnarray}
With this, Eq.~(23) of \cite{Golec-Biernat:1999qor} becomes 
\beq
x_{I\!\!P}F^D_{\{t,q\bar{q}g\}}(Q^2,\beta,x_{I\!\!P}) &=& \pi \beta  \int_\beta^1 \frac{d\xi}{\xi\beta'} \left( \left(1-\xi\right)^2+\xi^2\right)\int^{(1-\beta')Q^2} dk_\perp^2 \frac{\alpha_s}{2\pi}\ln \frac{(1-\beta')Q^2}{k_\perp^2} x'\frac{df_g(\beta', k_\perp;x_{I\!\!P})}{dY_{I\!\!P}} \nn 
&=&   \int_\beta^1 d\xi  \left( \left(1-\xi\right)^2+\xi^2\right)\int^{(1-\beta')Q^2} \frac{d^2k_\perp}{k_\perp^2}  \frac{\alpha_s}{2\pi^2} \int^{k_\perp^2} d^2k'_\perp x'\frac{df_g(\beta',k'_\perp;x_{I\!\!P})}{dY_{I\!\!P}}\,,
\eeq
where $\beta'=\beta/\xi$ and $x'=x_{I\!\!P}\beta'$. This can be recognized as the $g\to q$ DGLAP evolution of the collinear DPDF.   

We however note that (\ref{glueid}) is an {\it ad hoc}  prescription to correct for the inconsistent treatment of the quark and gluon saturation momenta $Q_s$ and $Q_{as}$ in \cite{Golec-Biernat:1999qor}. This introduces ambiguities.  For example, by normalizing ${\cal G}$ to match the integrated gluon distribution with the same saturation scale $Q_{as}=Q_s$ as for the quark distribution, the unitarity of the dipole S-matrix is spoiled. As a consequence, the numerical result for the integrated gluon DPDF differs from ours by a factor of 2. This can be seen by comparing the curves in our Fig.~5 and those in Fig.~3 of Ref.~\cite{Golec-Biernat:2001gyl} after adjusting the other normalization factors. 

\end{widetext}


\end{document}